\documentclass{article}
\usepackage{spconf,amsmath,booktabs,amssymb,graphicx,subfigure,threeparttable,multirow,float,subfig,bm,footnote,url}

\makeatletter 
\renewcommand{\@thesubfigure}{\hskip\subfiglabelskip}
\makeatother

\ninept

\title{Exploring wav2vec 2.0 on speaker verification and language identification}
\name{Zhiyun Fan$^{1,2}$, Meng Li$^1$, Shiyu Zhou$^1$, Bo Xu$^{1,2}$}
\address{
	$^1$Institute of Automation, Chinese Academy of Sciences, China\\
	$^2$School of Artificial Intelligence, University of Chinese Academy of Sciences, Beijing, China \\
	\{fanzhiyun2017, limeng, zhoushiyu2013, xubo\}@ia.ac.cn}
\begin{document}
%
\maketitle
\begin{abstract}
	Wav2vec 2.0 is a recently proposed self-supervised framework for speech representation learning. It follows a two-stage training process of pre-training and fine-tuning, and performs well in speech recognition tasks especially ultra-low resource cases. In this work, we attempt to extend the self-supervised framework to speaker verification and language identification. First, we use some preliminary experiments to indicate that wav2vec 2.0 can capture the information about the speaker and language. Then we demonstrate the effectiveness of wav2vec 2.0 on the two tasks respectively. For speaker verification, we obtain a new state-of-the-art result, Equal Error Rate (EER) of 3.61\% on the VoxCeleb1 dataset. For language identification, we obtain an EER of 12.02\% on the 1 second condition and an EER of 3.47\% on the full-length condition of the AP17-OLR dataset. Finally, we utilize one model to achieve the unified modeling by the multi-task learning for the two tasks.
\end{abstract}
\section{Introduction}
\label{sec:intro}
Recently, neural networks trained with a large amount of labeled data can meet most industrial needs in the field of speech processing \cite{li2019speechtransformer,DBLP:conf/interspeech/KannanDSWRWBCL19,xie2019utterance,nagrani2020voxceleb,padi2019attention,shen2019interactive}. However, purely supervised learning seems to be inconsistent with the mechanism of human learning. Early on in their lives, human infants learn language by watching and listening to adults around them, which resembles an unsupervised learning process. Later, they learn reading and writing, which seems to be a supervised learning process. To simulate the two-stage learning process, a lot of self-supervised frameworks are proposed \cite{DBLP:conf/naacl/DevlinCLT19,DBLP:conf/naacl/PetersNIGCLZ18,radford2018improving,DBLP:conf/icml/Henaff20,bachman2019learning,ravanelli2020multi}. 

In the field of speech processing, most self-supervised methods can be divided into two categories. One kind of method is conducted by the reconstruction loss, such as autoregressive predictive coding (APC) \cite{DBLP:conf/interspeech/ChungHTG19}, masked predictive coding (MPC) \cite{jiang2019improving} and so on. The other kind of method is conducted by contrastive predictive loss. The most representative work is the contrastive predictive coding (CPC) \cite{oord2018representation} and wav2vec \cite{DBLP:conf/interspeech/SchneiderBCA19}. The wav2vec 2.0 \cite{DBLP:conf/nips/BaevskiZMA20} used in this paper belongs to the latter category.
Most of these self-supervised pre-training methods are applied to speech recognition. However, there is almost no work on whether pre-training methods could work on the speaker verification (SV) or the language identification (LID). In this paper, we use the framework of wav2vec 2.0 \cite{DBLP:conf/nips/BaevskiZMA20} to explore this feasibility.

\begin{figure*}[ht]
	\centering
	\centerline{\includegraphics[width=12.0cm]{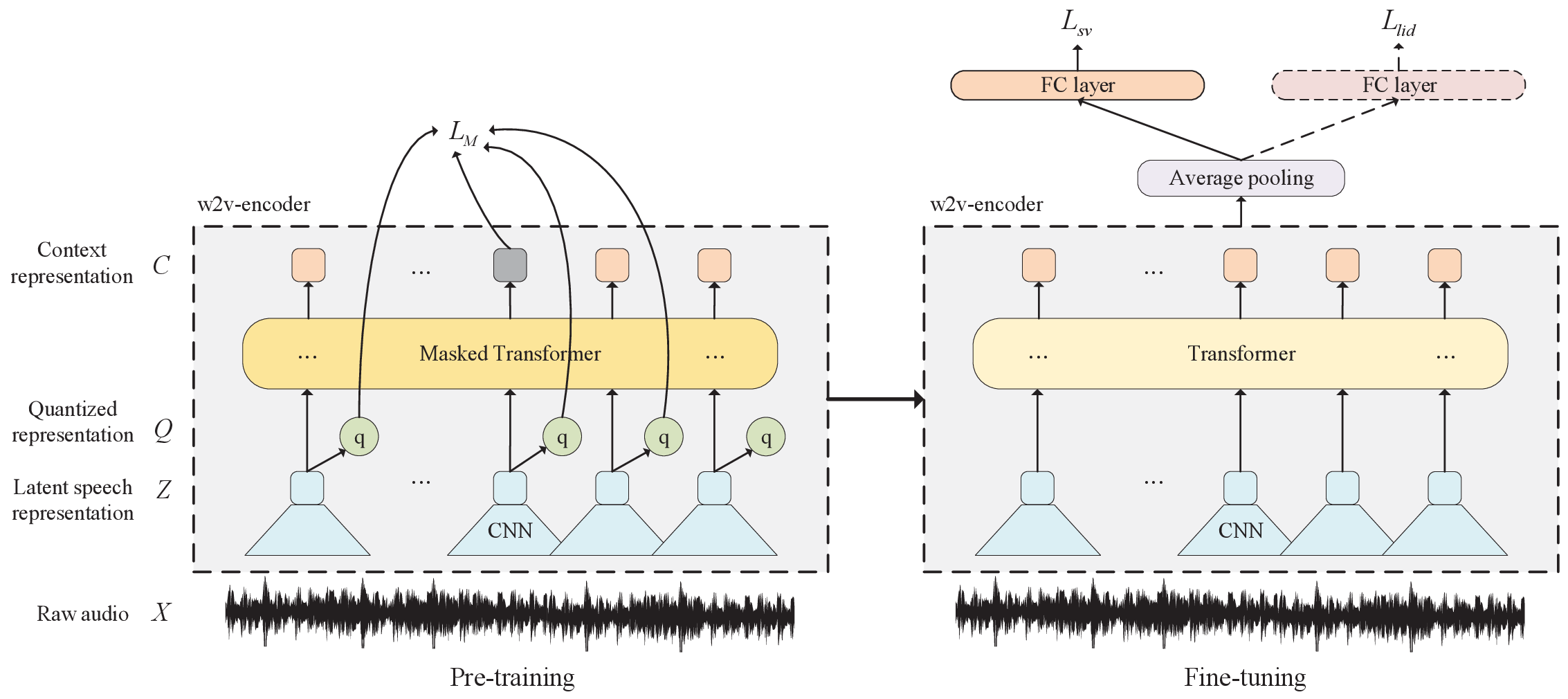}}
	\caption{An overview of the pre-training and fine-tuning. The models architecture used in pre-training stage and fine-tuning stage are identical, except the quantization modules and extra output layers.}
	\label{fig:F1}
	\hfill
\end{figure*}

We denote the model structure used in wav2vec 2.0 as w2v-encoder in this paper. It is illustrated in the dashed box of Fig.\,\ref{fig:F1}. It mainly consists of a convolutional neural network (CNN) encoder and a Transformer \cite{vaswani2017attention}. The CNN transfers raw waveform input to latent speech representations. They are fed to the Transformer after being masked and converted to context representations. A quantization module converts the latent speech representations to a discrete version which is used as the target. The whole model is trained to solve a contrastive task, which requires identifying the true quantized latent speech representations for a masked time step within a set of distractors \cite{DBLP:conf/nips/BaevskiZMA20}. After pre-training, Baevski et al. applied it to ultra-low resource speech recognition. Using only ten minutes labeled data, their approach achieved word error rate (WER) of 5.7/10.1\% on the clean/noisy test sets of Librispeech. The results demonstrate that the phoneme-related information is preserved during the pre-training of w2v-encoder and the drownstream task such as speech recognition can benefit a lot from it. Audio is a complex signal that contains not only phoneme-related information but also factors about speaker, language, environment, noise, etc. However, there is very little work on pre-training for the SV and the LID.

In this paper, we explore the effectiveness of self-supervised pre-training on the SV and the LID tasks. 
We utilize the pre-trained w2v-encoder to extract context representations, and use t-SNE \cite{maaten2008visualizing} tools to visualize them. We find that they have distinguishability among different speakers and languages even if pre-training of wav2vec 2.0 is problem-agnostic. Moreover, we find the lower layer has the stronger distinguishability. This distinguishability is exactly what the SV and the LID tasks need. It also verifies the feasibility of applying the self-supervised pre-training to the two tasks. Thus, we attempt to fine-tune the pre-trained model on these two downstream tasks respectively. For the SV task, we obtain an EER of 3.61\% on the test set of the VoxCeleb1 dataset. For the LID task, we obtain an EER of 12.02\% on the 1 second condition and 3.47\% on the full-length condition of the AP17-OLR dataset. Furthermore, in order to simplify the fine-tuning process and reduce model parameters, we utilize the multi-task learning to conduct the fine-tuning on the two tasks simultaneously.

\section{Method}
\label{sec:format}
In this section, we first review the pre-training of the wav2vec 2.0 \cite{DBLP:conf/nips/BaevskiZMA20}. Then we introduce how to apply the pre-trained model to downstream tasks. Fig.\,\ref{fig:F1} illustrates the pre-training and fine-tuning.
\subsection{Pre-training of wav2vec 2.0}
\label{ssec:subhead}
The left side of Fig.\,\ref{fig:F1} gives an illustration of the w2v-encoder and its pre-training. The main body of the model consists of a CNN-based feature encoder, a Transformer-based context network and a quantization module. The CNN encoder stacks seven blocks, and in each block the temporal convolutions followed by a GELU activation function \cite{hendrycks2016gaussian} have 512 channels with strides $(5,2,2,2,2,2,2)$ and kernel widths $(10,3,3,3,3,2,2)$. The CNN encoder maps the raw audio $X$ into latent speech representations $Z$. 

The context network stacks 12 Transformer blocks with model dimension $768$, inner dimension $3,072$, and $8$ attention heads. Before sending $Z$ into the context network, all time steps of $Z$ are randomly sampled as starting indices with $p=0.065$, and $M=10$ consecutive time steps from every sampled index are masked. Then the relative positional embedding is added to the masked representations. The Transformer contextualizes the masked representations and generates context representations $C$. 

The quantization module is used to discretize latent speech representations $Z$ into $Q$. There are $G=2$ codebooks in the quantization module. Each of them contains $V=320$ entries with a size of $128$. The quantization module firstly maps the $Z$ to logits $l \in \mathbb{R}^{G\times V}$. Then the gumbel softmax \cite{gumbel1954statistical} is used to choose one entry from each codebook in a fully differentiable way. All the entries selected are concatenated to resulting vectors $[e_1;e_2;...;e_G]$, which are linearly mapped to $q$. The loss function is as follows:
\begin{equation}
	L=L_M+\alpha L_D+\beta L_F
	\label{eq1}
\end{equation}
\begin{equation}
	L_M=-{\rm log}\frac{{\rm exp}(sim(c_t, q_t))/k}{\sum_{\tilde{q}\sim Q_t}{\rm exp}(sim(c_t, \tilde{q}))/k}
	\label{eq2}
\end{equation}
\begin{equation}
	L_D=\frac{1}{GV}\sum_{g=1}^{G}\sum_{v=1}^{V}\bar{p}_{g,v}{\rm log}\bar{p}_{g,v}
	\label{eq3}
\end{equation}
\begin{equation}
	\bar{p}={\rm GumbelSoftmax}(\bar{l})
	\label{eq4}
\end{equation}

The loss is the weighted sum of three terms. In the Eq.\,\ref{eq1}, $L_F$ is a $L2$ penalty. The weight $\beta$ is set to 10. The $L_M$ is the contrastive loss to make the model distinguish true discrete representations from latent distractors $\tilde{q}$. The distractors are uniformly sampled from other masked time steps of the same utterance. In Eq.\,\ref{eq2}, the $sim$ represents cosine similarity, and the $Q_t$ includes $q_t$ and $K=100$ distractors, and the temperature $k$ is set to $0.1$. The $L_D$ is the diversity loss designed to increase the use of the quantized codebook representations. The $\alpha$ in Eq.\,\ref{eq1} is set to 0.1. The $\bar{l}$ in Eq.\,\ref{eq4} represents the average of logits $l$ across utterances in a batch.

The pre-training process is optimized with Adam \cite{kingma2014adam}. During the first 8\% of the updates, the learning rate warms up to a peak of $5\times10^{-3}$, and then it decays linearly. For more details about the pre-training of wav2vec 2.0, we refer readers to \cite{DBLP:conf/nips/BaevskiZMA20}.
\subsection{Fine-tuning}
\label{Fine-tuning}
Before the post-training, we add an average pooling layer and a fully connected layer on the top of w2v-encoder. The average pooling layer converts the frame-level context representations given by w2v-encoder into sentence-level representations, and the fully connected layer classifies each sentence into some speaker or some language.

The newly added fully connected layer is randomly initialized, and w2v-encoder is initialized with the base model released by Baevski et al \footnote{\url{https://github.com/pytorch/fairseq/blob/master/examples/wav2vec/}}. The cross-entropy criteria is employed as the loss function for the classification of speakers or languages. Specially, for the training of speaker classification, AM-softmax \cite{wang2018additive} is used to increase the discrimination of the learned embedding to the speaker.

In the multi-task fine-tuning, we add a pooling layer and two parallel fully connected layers to predict the speaker and language respectively. The training loss is obtained by the weighted sum of the losses of these two tasks. The $L_{sv}$ and $L_{lid}$ in Eq. \ref{eq5} represent the CE loss of the SV and the LID tasks respectively.
\begin{equation}
	L_{mul}=\lambda L_{sv}+(1-\lambda) L_{lid}
	\label{eq5}
\end{equation}

Due to the problem of unbalanced data volume in the datasets of the speakers and languages, the batch is generated by sampling from two datasets with equal probability to ensure the data used in the training process is balanced. In addition, the two tasks also have the problem of inconsistent convergence speed. We mitigate this issue by adjusting the weight of the loss of the two tasks through the development set.
\section{Experiments}
\label{sec:pagestyle}
Various informative factors are mixed in speech signals, including semantics, speaker, emotion, channel, background noise, etc. Baevski et al. have shown that the representations underlying pre-trained w2v-encoder can capture the linguistic factors. It remains unclear whether the problem-agnostic pre-training of wav2vec 2.0 can learn about any other factors. In the experiment part, we take speaker and language factors as examples to explore this question, and try to apply wav2vec 2.0 to the SV and the LID tasks. 
\subsection{Datasets}
\label{datasets}
VoxCeleb1 \cite{nagrani2017voxceleb} and AP17-OLR \cite{DBLP:journals/corr/abs-1806-00616} datasets are used in our experiments for the SV and the LID respectively.

\textbf{Speaker verification dataset}: VoxCeleb1 \cite{nagrani2017voxceleb} contains over 100,000 utterances from 1,251 celebrities. It can be used for both speaker identification and verification. We use the VoxCeleb1 to conduct the SV task. And the consine distance is used to calculate the similarity score. The data split of the VoxCeleb1 dataset for verification is listed in Table \ref{VoxCeleb1}.
\begin{table}[htbp]
	\caption{Data split of the VoxCeleb1 dataset for verification.}
	\vspace{-6.5pt}
	\label{VoxCeleb1}
	\centering
	\setlength{\tabcolsep}{5.0mm}{
		\begin{tabular}{cccc}
			\toprule
			& \textbf{Train} & \textbf{Valid} & \textbf{Test}\\
			\hline
			\#speakers & 1211   &1145  &40  \\
			\#Utterances & 143642 & 5000  & 4874 \\
			Dur(hrs.) & 329.06 & 11.34    &11.20\\
			\bottomrule
		\end{tabular}}
	\end{table}
	
\textbf{Language identification dataset}: AP17-OLR \cite{DBLP:journals/corr/abs-1806-00616} consists of 10 different languages (Mandarin, Cantonese, Indonesian, Japanese, Russian, Korean, Vietnamese, Kazakh, Tibetan and Uyghur). The duration of training data for each language is about 10 hours with the speech sampled at 16 kHz. The test set contains three subsets with different durations (1 second, 3 second, and full length). These subsets respectively contain 17964, 16404 and 17964 utterances. 
\subsection{Model description}
	\label{datasets}
	In the experiments, we utilize the base model released by Baevski et al. and three models fine-tuned by us. For simplicity, we use some symbols to represent them, and the explanations are as follows:
	\begin{itemize}
		\item \textbf{M-nofinetune}: the base model pre-trained on the Librispeech corpus \cite{panayotov2015librispeech}.
		\item \textbf{M-sv}: We fine-tune M-nofinetune on the VoxCeleb1 dataset for speaker verification.
		\item \textbf{M-lid}: We fine-tune M-nofinetune on the AP17-OLR dataset for language identification. 
		\item \textbf{M-multi}: We fine-tune M-nofinetune on the AP17-OLR and VoxCeleb1 dataset simultaneously in a multi-task form.		
	\end{itemize}
	\subsection{Feasibility analysis}
	\label{Preliminary analysis}
	\begin{figure}[!ht]
		\subfigure[(a1) layer12]{
			\begin{minipage}[t]{0.5\linewidth}
				\centering
				\centerline{\includegraphics[width=3.3cm]{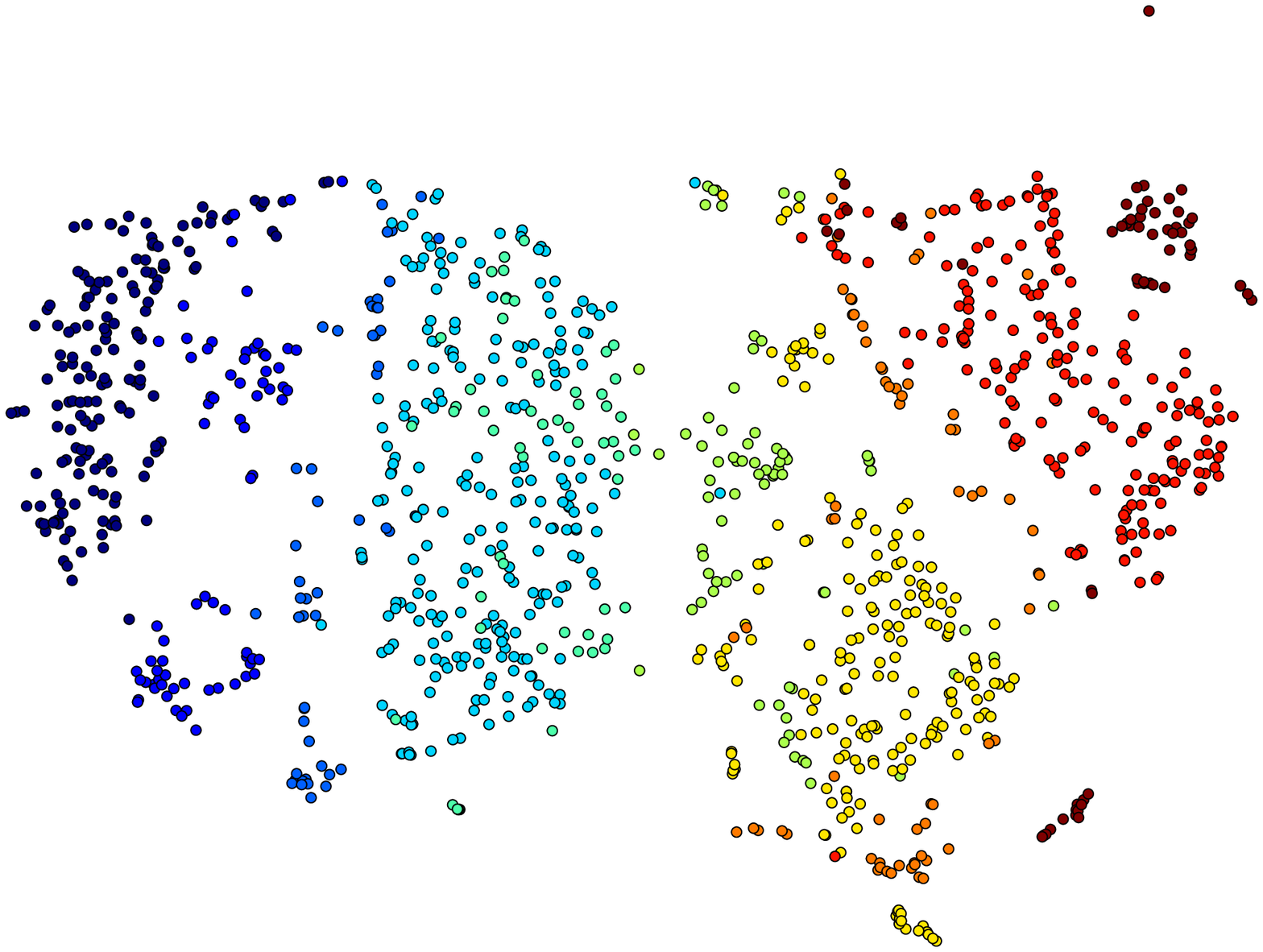}}
			\end{minipage}}
			\subfigure[(b1) layer12]{
				\begin{minipage}[t]{0.5\linewidth}
					\centering
					\centerline{\includegraphics[width=3.3cm]{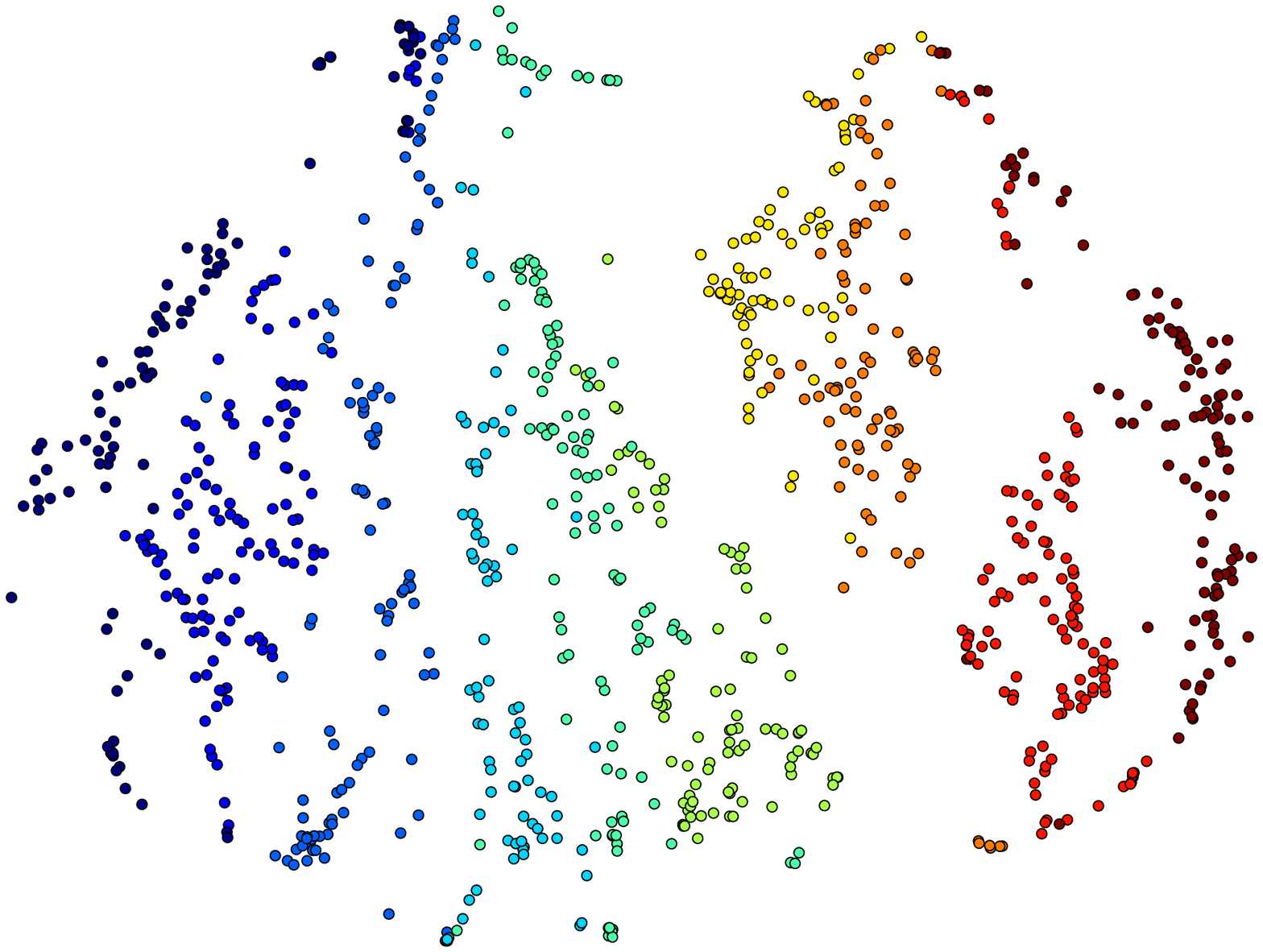}}
				\end{minipage}}
				
				\subfigure[(a2) layer6]{
					\begin{minipage}[t]{0.5\linewidth}
						\centering
						\centerline{\includegraphics[width=3.3cm]{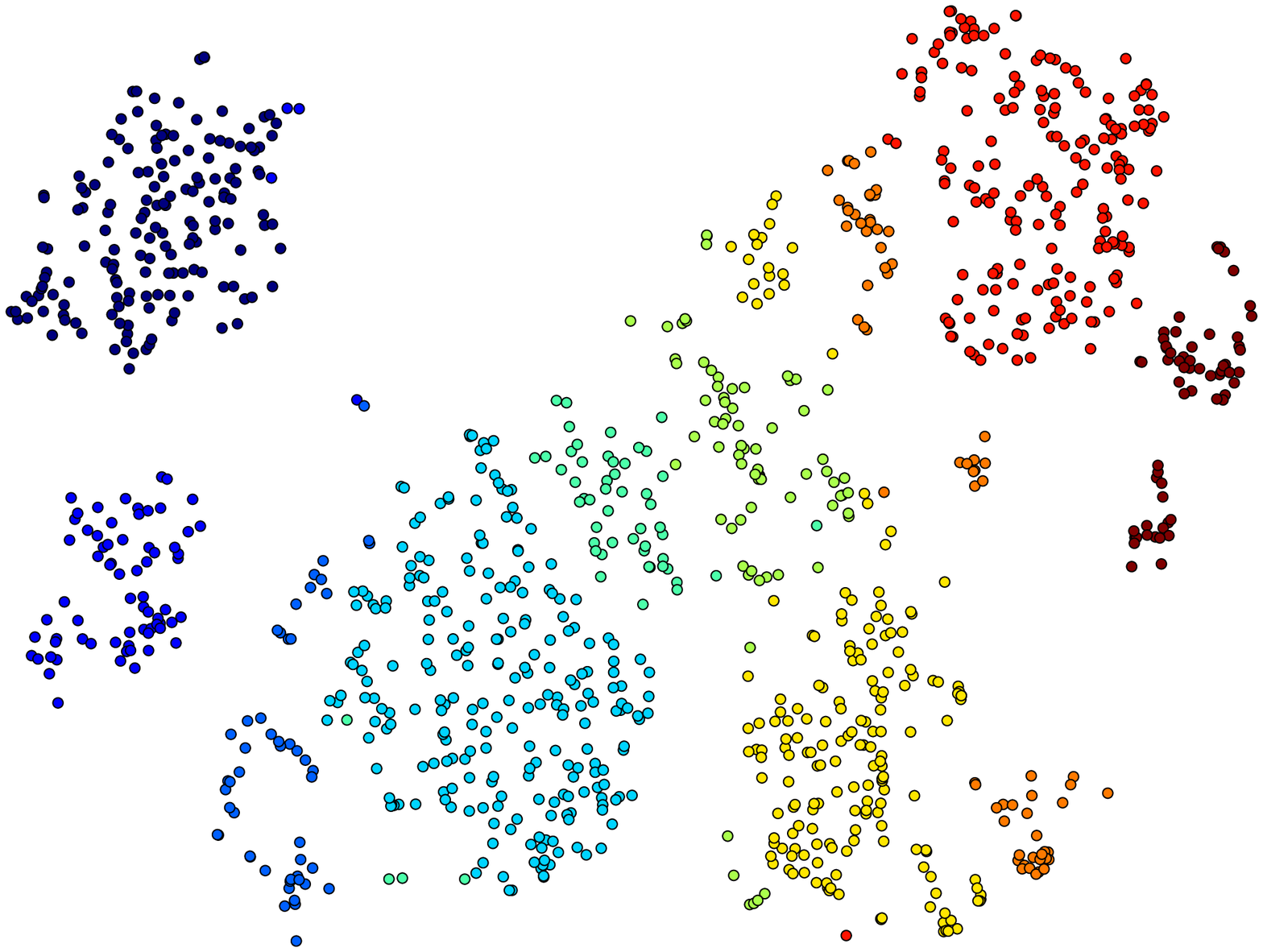}}
					\end{minipage}}
					\subfigure[(b2) layer6]{
						\begin{minipage}[t]{0.5\linewidth}
							\centering
							\centerline{\includegraphics[width=3.3cm]{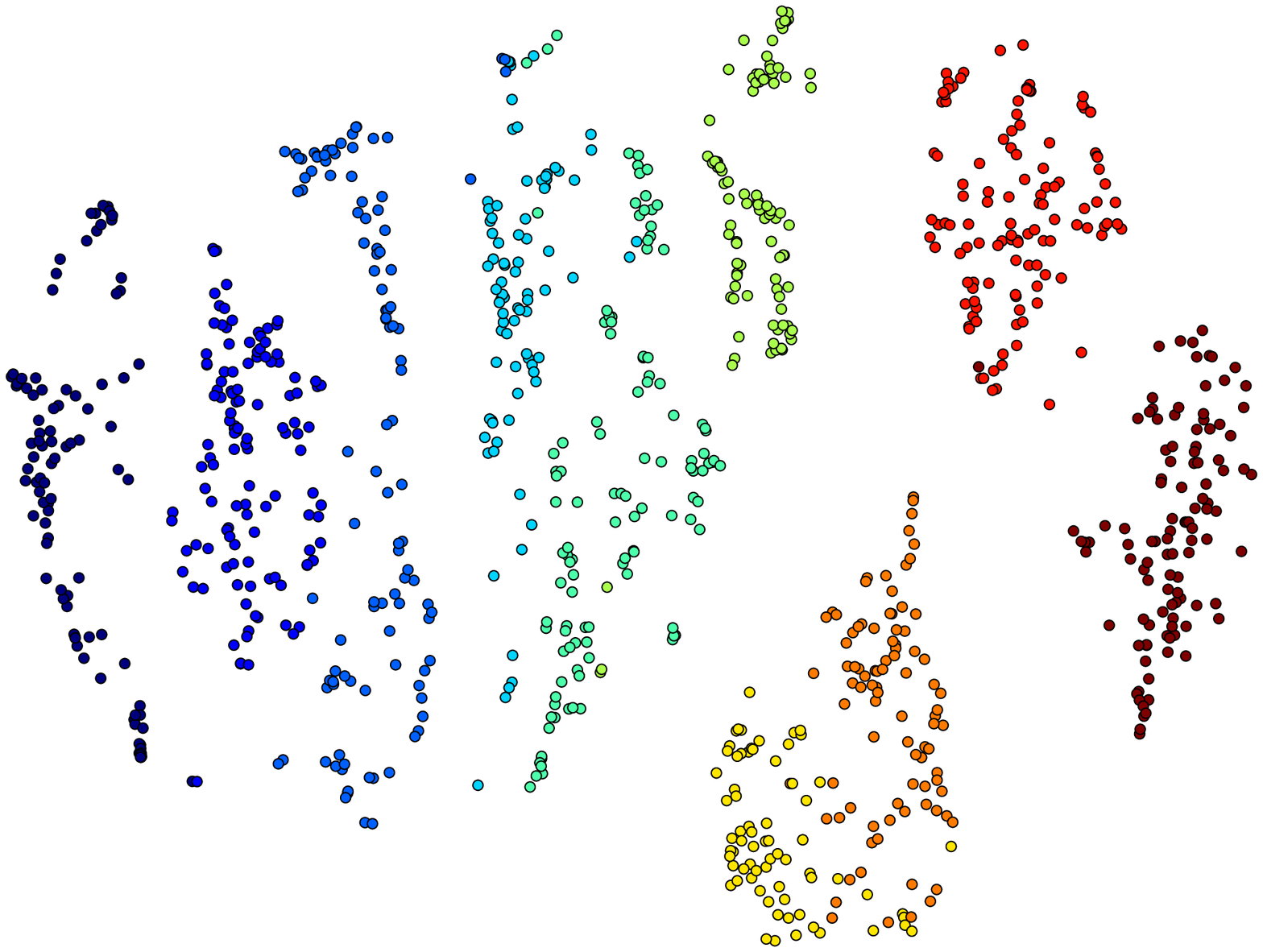}}
						\end{minipage}}

						\subfigure[(a3) layer1]{
							\begin{minipage}[t]{0.5\linewidth}
								\centering
								\centerline{\includegraphics[width=3.3cm]{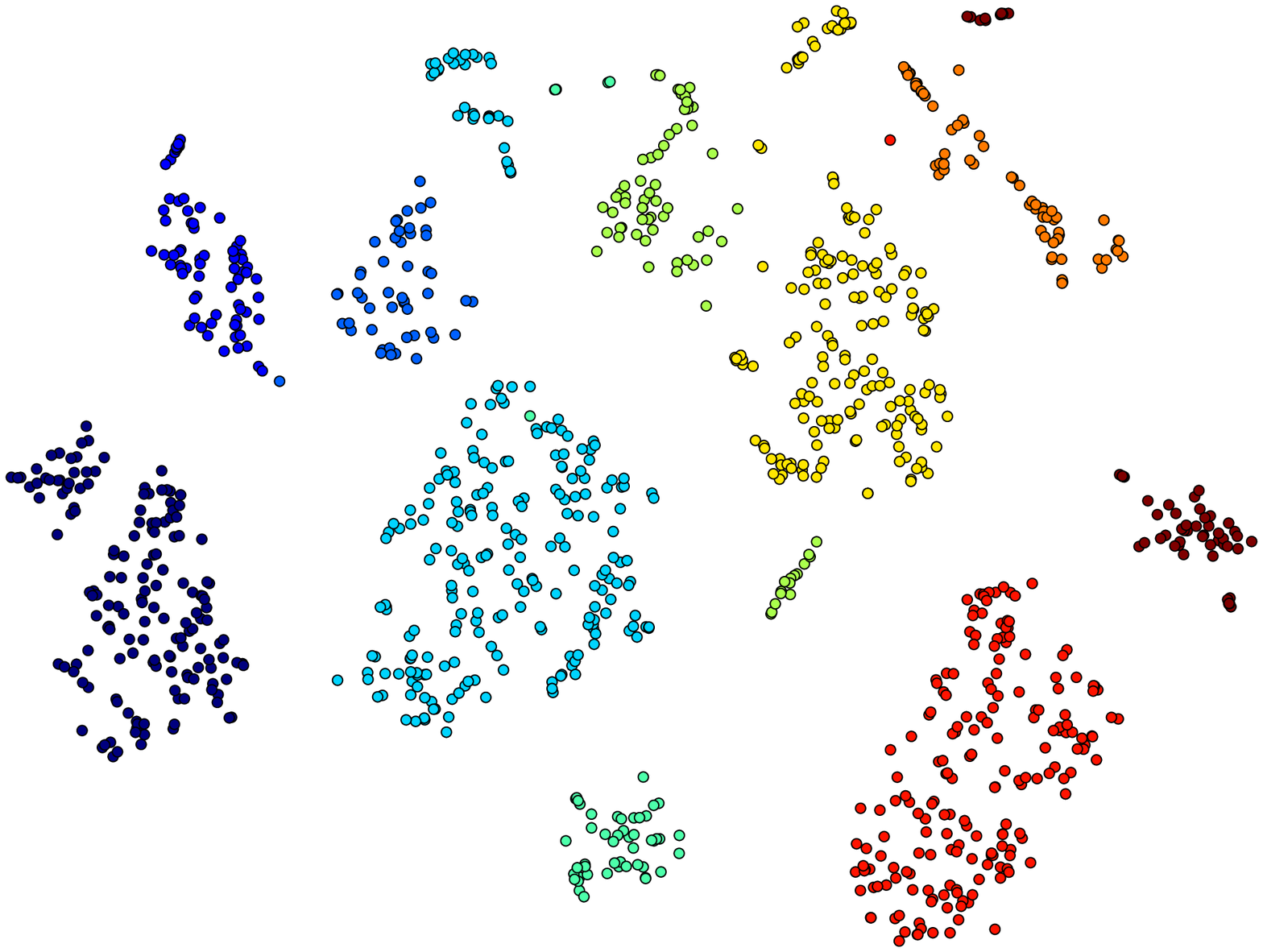}}
							\end{minipage}}
							\subfigure[(b3) layer1]{
								\begin{minipage}[t]{0.5\linewidth}
									\centering
									\centerline{\includegraphics[width=3.3cm]{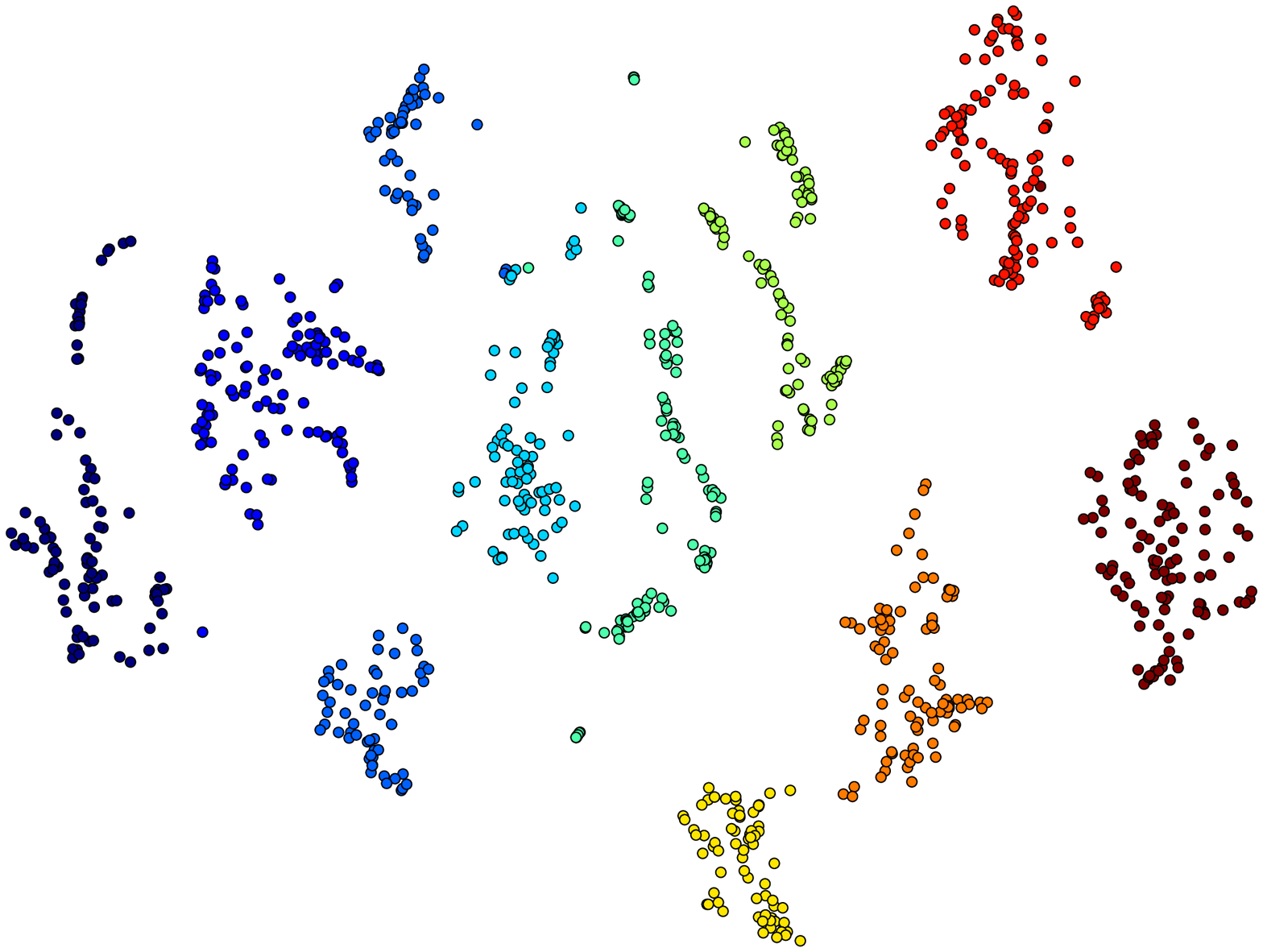}}
								\end{minipage}}
								\caption{2D t-SNE plot of representations extracted from the bottom layer (layer1), the middle layer (layer6), and the high layer (layer12) of the Transformer of M-nofinetune. The left column is the clustering of the representations of 10 speakers in the test set of VoxCeleb1 extracted by M-nofinetune, and each color represents a speaker. The right column is the clustering of the representations of 1000 samples in the test set of the AP17-OLR extracted by M-nofinetune, and each color represents a language.}
								\label{fig:res}
							\end{figure}
In this section, we explore whether the speaker and language factors are retained during the pre-training of wav2vec 2.0. It determines whether the pre-training method can be used for these two tasks. 

We directly extract context representations from the test set of AP17-OLR and VoxCeleb1 with the M-nofinetune model. Then we visualize the context representations by t-SNE \cite{maaten2008visualizing}, a nonlinear dimensionality reduction algorithm for visualizing high-dimensional data. The results are shown in Fig.\,\ref{fig:res}. The left three images are the visualization results of the features from the three layers of the Transformer. Different colors represent different speakers. It is not difficult to find that all the three layers have certain speaker distinguishability, and this distinguishability is more obvious at the bottom of the Transformer. In the three images on the right, different colors represent different languages. It can also be found that these features are distinguished by languages, and the lower the layer, the stronger the distinction. The phenomena presented in Fig.\,\ref{fig:res} show that the model pre-trained by wav2vec 2.0 can effectively extract the characteristics of the speaker and language of the speech.
							
We further quantify this claim by performing the SV and the LID with a simple fully connected layer. The pre-trained w2v-encoder (M-nofinetune) acts as a feature extractor. The fully connected layer is optimized to distinguish the 10 languages or 1211 speakers for the two tasks respectively. The test results are listed in Table \ref{t1}.
							\begin{table}[htbp]
								\caption{The EER (\%) results of using context representations extracted by M-nofinetune to finish the SV and the LID.}
								\vspace{-6.5pt}
								\label{t1}
								\centering
								\setlength{\tabcolsep}{5.0mm}{
									\begin{tabular}{ccc}
										\toprule
										\textbf{Model} & \textbf{SV} & \textbf{LID} \\
										\hline
										random & 47.93   & 50.05  \\
										M-nofinetuning & 15.62 & 42.34  \\
										\bottomrule
									\end{tabular}}
								\end{table}
								
The $random$ results are evaluated on the randomly initialized w2v-encoder. The comparison of these two results in Table \ref{t1} further illustrates that the model pre-trained by wav2vec 2.0 can extract speaker and language-related characteristics, which provides a basis for the application of wav2vec 2.0 to the SV and the LID.
								%
\subsection{Speaker verification}
\label{ssec:subhead}
From the experiments in the previous section, we can see that the pre-trained w2v-encoder, M-nofinetune, can extract features that contain a certain speaker distinguishability. This kind of speaker distinguishing learning is exactly required in the SV task. Then we attempt to fine-tune the pre-trained w2v-encoder, M-nofinetune, to finish the SV task. We initialize the w2v-encoder with M-nofinetune, and add a randomly initialized fully connected layer on the top of it to predict speakers. The fine-tuning is conducted on the VoxCeleb1 dataset \cite{nagrani2017voxceleb}. All parameters are adjustable during fine-tuning. However, at the first $10000$ steps, the w2v-encoder is frozen. We optimize the model with Adam, the learning rate warms up to $5 \times 10^{-3}$ during the first $6000$ steps, and then it decays linearly during the remaining $7000$ steps. 
	\begin{table}[htbp]
		\caption{Comparison with the previously published EER (\%) results on Voxceleb1 dataset.}
			\vspace{-6.5pt}
			\label{results-vox1}
			\centering
		\setlength{\tabcolsep}{7mm}{
			\begin{tabular}{cc}
			\toprule
			\textbf{Model}& \textbf{EER}\\
			\hline
		    I-vectors + PLDA \cite{nagrani2017voxceleb} &  8.8   \\
			CNN + Embedding \cite{nagrani2017voxceleb} &    7.8    \\
			LDE-ASoftmax \cite{cai2018exploring} & 4.41 \\
			attentive statistics \cite{okabe2018attentive} & 3.85 \\
			no pre-training & 24.28 \\
			M-sv &  $\bm{3.61}$    \\
		\bottomrule
				\end{tabular}}
			\end{table}
									
The $no\ pre$-$training$ in Table \ref{results-vox1} represents that using VoxCeleb1 to train the w2v-encoder added a fully connected layer without pre-training. Our fine-tuning model, M-sv, outperforms the $no\ pre$-$training$ result by a significant margin (EER of $3.61\%$ vs $24.28\%$). The gap between them illustrates the benefits of pre-training. Moreover, our model outperforms all baselines in Table \ref{results-vox1}, and obtains new state-of-the-art results on the VoxCeleb1 dataset. It means the pre-training of wav2vec 2.0 is useful to the SV task and can work well without any task-specific adjustment of model structure.
\subsection{Language identification}
	\label{ssec:subhead}
Although Baevski et al. only used English data during the pre-training of M-nofinetune, it can be seen from the visualization results in section \ref{Preliminary analysis} that the features extracted by M-nofinetune still retain the distinction of language. It means that the model obtained by this pre-training method may be useful to the language identification system. Similarly, we add a fully connected layer on top of the w2v-encoder to predict language. We initialize w2v-encoder with M-nofinetune and randomly initialize the extra fully connected layer. Then the whole model is fine-tuned on the AP17-OLR dataset \cite{DBLP:journals/corr/abs-1806-00616}. We optimize the model with Adam, the learning rate warms up to $5 \times 10^{-3}$ during the first $5000$ steps, and then it decays linearly during the remaining $8000$ steps. The parameters of the w2v-encoder part are frozen at the first $5000$ steps. After training, we test on the model, which obtains the best performance on the development set. 
									\begin{table}[htbp]
										\begin{threeparttable}
											\caption{Comparison with the previously published $C_{avg}$ and EER (\%) results on AP-17 dataset.}
											\vspace{-6.5pt}
											\label{results:ap17}
											\centering
											\setlength{\tabcolsep}{1.2mm}{
												\begin{tabular}{ccccc}
													\toprule
													\multirow{2}{*}{\textbf{Model}} & 
													\multicolumn{2}{c}{\textbf{1 second}}&\multicolumn{2}{c}{\textbf{Full-Length}}\cr
													\cmidrule(lr){2-3} \cmidrule(lr){4-5}
													& \bm{$C_{avg}$} & \textbf{EER} & \bm{$C_{avg}$} & \textbf{EER} \cr
													\midrule
													i-vector + PLDA\cite{DBLP:journals/corr/abs-1806-00616} &  0.1746  & 17.51 & 0.0596 & 5.86   \\
													TDNN \cite{DBLP:journals/corr/abs-1806-00616} & 0.1282 & 14.04 & 0.1034 & 11.31 \\
													TSM-DNN-BN-LSTM \cite{ma2018short} & $\bm{0.067}$ & $\bm{6.95}$ & $\bm{0.007}$ & $\bm{0.86}$ \\
													no pre-training & 0.2813 & 29.25 &  0.1254 & 13.83  \\
													M-lid & 0.1158 & 12.02 & 0.0310 & 3.47  \\
													\bottomrule
												\end{tabular}}
											\end{threeparttable}
										\end{table}	
										
The first two rows in Table \ref{results:ap17} are the two baselines released by the organizer of the AP17-OLR challenge. The $TSM$-$DNN$-$BN$-$LSTM$ \cite{ma2018short} is one of the best models on this benchmark. The $no\ pre$-$training$ in Table \ref{results:ap17} means that the fine-tuning starts from scratch on the AP17-OLR dataset. The M-lid, which is fine-tuned from the pre-trained M-nofinetune, outperforms the $no\ pre$-$training$ result by a large margin on both the 1 second condition and the full-length condition. The gap between them illustrates the benefits brought by pre-training to the LID task. Compared with baselines released by the organizer, M-lid shows a clear performance advantage on the two test conditions. However, it is far from the best results. It means that wav2vec 2.0 is useful to the LID task. However, its effectiveness on the LID task is not good as the SV task. We consider that the use of multiple languages during pre-training (not just English) can mitigate this issue. In addition, we find that the performance of the $no\ pre$-$training$ is influenced by overfitting seriously. This problem is obviously alleviated during the fine-tuning of M-lid, which benefits from pre-training.									
\subsection{Multi-task system}
\label{ssec:subhead}
The parameters of the w2v-encoder have reached $94$M. Fine-tuning two models for the SV and the LID tasks independently will take up a lot of resources. Hence, we try to use one model to finish these two tasks simultaneously. On the top of the w2v-encoder we connect two fully connected layers in parallel to predict the speaker and language respectively. We follow the experiment settings described in section \ref{Fine-tuning}.
The $\lambda$ in Eq.\,\ref{eq5} is set to $0.7$.
										\begin{table}[htbp]
											\caption{Performance of single-task model and multi-task model on the VoxCeleb1 and the AP17-OLR dataset in terms of EER(\%).}
											\vspace{-6.5pt}
											\label{results-mul}
											\centering
											\setlength{\tabcolsep}{7mm}{
												\begin{tabular}{ccc}
													\toprule
													\textbf{Model}& \textbf{SV} & \textbf{LID} \\
													\hline
													M-sv &  3.61  & - \\
													M-lid &    - &  3.47  \\
													M-multi &  4.18  &  4.88  \\
													\bottomrule
												\end{tabular}}
											\end{table}
											
Results in Table \ref{results-mul} show that compared with single-task training, although the performance of multi-task form is a bit reduced, it achieves good results with fewer parameters on the SV and the LID tasks. It shows that the pre-training of wav2vec 2.0 can be combined with multi-task learning to achieve unified modeling of the two tasks. This greatly simplifies the use of pre-trained model and can save a lot of time spent on fine-tuning to each task. In addition, it can reduce the demand for storage.
\section{Conclusion}
\label{sec:typestyle}
In this paper, we explore the application of wav2vec 2.0 on speaker verification and language identification. First of all, through some preliminary experiments and visualization methods, we find that the features extracted by the pre-trained w2v-encoder have the distinction between speakers and languages, and this distinction is more obvious in lower layers. This illustrates the feasibility of using the pre-trained model for the SV and the LID tasks. Then we verify the effectiveness of the pre-trained model on the two tasks and obtain competitive results on the VoxCeleb1 and the AP17-OLR datasets. Finally, in order to simplify the fine-tuning process on multiple tasks and reduce parameters, we use a multi-task learning mechanism, so as to realize the unified modeling for the SV and the LID. In future work, we are planning to extend wav2vec 2.0 to more speech processing tasks with the multi-task learning.

\vfill\pagebreak

\bibliographystyle{IEEEbib}
\bibliography{strings,refs}

\end{document}